\documentclass{elsart}

\usepackage{amssymb,amsmath,mathrsfs}
\usepackage{feynmp}
\usepackage{graphicx}

\newcommand{\Lag}{\mathscr{L}}
\newcommand{\dyad}[1]{\ensuremath{\overset{\leftrightarrow}{#1}}}
\newcommand{\funcd}[2]{\frac{\delta #1}{\delta #2}}
\renewcommand{\i}{\text{i}}
\DeclareMathOperator{\Tr}{Tr}
\def\Im{\mathrm {Im}\,}
\begin{document}

\begin{frontmatter}
  
\title{Finite pion width effects on the rho--meson} 

\author{Hendrik van Hees\thanksref{hvh}} and \author{J\"o{}rn
  Knoll\thanksref{jk}}
  
\thanks[hvh]{e-mail:h.vanhees@gsi.de}

\thanks[jk]{e-mail:j.knoll@gsi.de}

\address{Gesellschaft f\"u{}r Schwerionenforschung \\ Planckstr. 1 \\
  64291 Darmstadt}

\date{10 July 2000}

\begin{abstract}\noindent
  We study the influence of the finite damping width of pions on the
  in--medium properties of the $\rho$--meson in an interacting meson
  gas model at finite temperature. Using vector dominance also
  implications on the resulting dilepton spectra from the decay of the
  $\rho$--meson are presented. A set of coupled Dyson equations with
  self--energies up to the sunset diagram level is solved self
  consistently. Following a $\Phi$--derivable scheme the self--energies
  are dynamically determined by the self--consistent propagators. Some
  problems concerning the self--consistent treatment of vector or gauge
  bosons on the propagator level, in particular, if coupled to
  currents arising from particles with a sizable damping width, are
  discussed.
\end{abstract}


\end{frontmatter}

\section{Introduction and summary}

The question how a dense hadronic medium changes the properties of
vector mesons compared to their free space characteristics has
attracted much attention in recent times. Experimentally this question
is studied in measurements of the dilepton production rates in heavy
ion collisions. Recent experiments by the CERES and DLS collaborations
\cite{ceres95,ceres95-2,dls97} show that the low lepton pair mass
spectrum is significantly enhanced in the range between
$300\,\text{MeV}$ and $600\,\text{MeV}$ compared to the yield that one
expects from the corresponding rates in pp--collisions.

>From the theoretical side various mechanisms are proposed to explain
these influences of a hadronic matter surrounding on the spectral
properties of vector
mesons\cite{gale-kap90,rw99,hfn93,rcw97,rw99b,ubrw98,rubw98,rg99}.
With the upgrade of CERES and the new dilepton project HADES at GSI a
more precise view on the spectral information of vector mesons is
expected. Still in most of theoretical investigations the damping
width gained by stable particles due to collisions in dense matter is
either ignored or treated within an extended perturbation theory
picture\cite{ubw99}. In this contribution we study the in--medium
properties of the $\rho$--meson due to the damping width of pions in a
dense meson gas within a self--consistent scheme.

The field theoretical model, which is inspired from vector meson
dominance theories\cite{klz67}, is discussed in section 2. Thereby the
finite pion width is modelled with a four--pion self--interaction in
order to keep the investigation as simple as possible. The coupling
strength is adjusted as to produce a pion damping--width of reasonable
strength, such as to simulate the width a pion would obtain in a
baryon rich environment due to the strong coupling to baryonic
resonance channels, like the $\Delta$--resonance. The self--consistent
equations of motion are derived from Baym's
$\Phi$--functional\cite{lw60,baym62,cjt74} and the problem of
renormalisation is left aside by taking into account only the
imaginary parts of the self--energies but keeping the normalisation of
the spectral function fixed.

This self--consistent treatment described in section 3 respects the
conservation laws for the expectation values of conserved currents and
at the same time ensures the dynamical as well as the thermodynamical
consistency of the scheme. Especially the effects of bremsstrahlung
and annihilation processes are taken into account consistently.

Finally in section 4 we discuss the principal problems with the
treatment of vector mesons in such a scheme which are mainly due to
the fact that within the $\Phi$--functional formalism the vertex
corrections necessary to ensure the Ward--Takahashi identities for the
propagator are ignored. In our model calculations we work around this
problem by projecting onto the transverse part of the propagators such
that the errors of this shortcoming can be expected to be small. In
the appendix we elaborate on some details of this projection method.
 
\section{The model}

In order to isolate the pion width effects we consider a purely mesonic model
system consisting of charged pions, neutral $\rho$--mesons, and also the
chiral partner of the $\rho$--, the $a_1$--meson, with the interaction
Lagrangian
\begin{equation}\label{Lint}
\Lag^{\text{int}}=g_{\rho\pi\pi}\rho_{\mu} \pi^* \dyad{\partial}{}^{\mu} 
\pi+g_{\pi\rho a_1}\pi\rho_{\mu} a_1^{\mu}+\frac{g_{\pi 4}}{8}
(\pi^* \pi)^2 + cc.
\end{equation}
We do not explicitely write down the free Lagrangian, but we like to mention
that we consider the $\rho$-meson as a gauge particle.  The first two coupling
constants are adjusted to provide the corresponding vacuum widths of the
$\rho$-- and $a_1$--meson at the nominal masses of $770\, \text{MeV}$ and
$1200\,\text{MeV}$ and widths of $\Gamma_{\rho}=150\,\text{MeV}$ and
$\Gamma_{a_1}=400\,\text{MeV}$, respectively. The four--$\pi$ interaction is
used as a tool to furnish additional collisions among the pions. The idea of
this term is to provide pion damping widths of about $50\,\text{MeV}$ or more
as they would occur due to the strong coupling to the $NN^{-1}$ and $\Delta
N^{-1}$ channels in an environment at finite baryon density.

The $\Phi$--functional method originally invented by Baym\cite{baym62}
provides a self--consistent scheme applicable even in the case of broad
resonances. It is based on a resummation for the partition sum
\cite{lw60,cjt74}.  Its two particle irreducible part $\Phi\{G\}$
generates the irreducible self--energy $\Sigma(x,y)$ via a functional
variation with respect to the propagator $G(y,x)$, i.e.
\begin{equation}
\label{varphi}
-\i \Sigma (x,y) =\funcd{\i\Phi}{\i G(y,x)}.  
\end{equation}
Thereby $\Phi$, constructed from two-particle irreducible closed
diagrams of the Lagrangian (\ref{Lint}) solely depends on fully
resummed, i.e.  self--consistently generated propagators $G(x,y)$. In
graphical terms, the variation (\ref{varphi}) with respect to $G$ is
realized by opening a propagator line in all diagrams of $\Phi$.
Further details and the extension to include classical fields or
condensates into the scheme are given in ref.~\cite{kv97}.

Truncating $\Phi$ to a limited subset of diagrams, while preserving the
variational relation (\ref{varphi}) between $\Phi^{\text{(appr.)}}$ and
$\Sigma^{\text{(appr.)}}(x,y)$ defines an approximation with built--in
consistency.  Baym\cite{baym62} showed that such a Dyson resummation scheme is
conserving at the expectation value level of conserved currents related to
global symmetries, realised as a linear representation of the corresponding
group, of the original theory, that its physical processes fulfil detailed
balance and unitarity and that at the same time the scheme is
thermodynamically consistent. However symmetries and conservation laws may no
longer be maintained on the correlator level, a draw--back that will lead to
problems for the self--consistent treatment of vector and gauge particles on
the propagator level, as discussed in sect. 3.

Interested in the effects arising from the damping width of the
particles we discard all changes in the real part of the self
energies, keeping however the sum--rule of the spectral functions
normalised. In this way we avoid renormalisation problems which
require a temperature independent subtraction scheme. The latter will
be discussed in detail in a forthcoming paper \cite{vHJK00}.
Neglecting changes in real parts of the self--energies also entitles to
drop tadpole contributions. The treatment of the tensor structure of
the $\rho$-- and $a_1$--polarisation tensors is discussed in sect. 3.
Here we first discuss the results of the self--consistent calculations.

For our model Lagrangian (\ref{Lint}) and neglecting tadpole
contributions one obtains the following diagrams for $\Phi$ at the
two--point level which generate the subsequently given three self
energies $\Pi_{\rho}$, $\Pi_{a_1}$ and $\Sigma_{\pi}$

\def\fmfsdot#1{\fmfv{decor.shape=circle,decor.filled=full,decor.size=1.2thick}
{#1}}
\def\GPhipirhopi{
\parbox{12mm}{
\begin{fmfgraph*}(12,0)
\fmfpen{thick}
\fmfleft{l}
\fmfright{r}
\fmfforce{(0.0w,0.5h)}{l}
\fmfforce{(1.0w,0.5h)}{r}
\fmf{gluon,left=.1,label=$\rho$,l.d=7.5,label.side=right}{r,l}
\fmf{plain,left=.9,tension=.5,label=$\pi$}{l,r}
\fmf{plain,left=.9,tension=.5,label=$\pi$,l.side=right}{r,l}
\fmfsdot{l,r}
\end{fmfgraph*}
}}
\def\GPhipirhoa{
\parbox{12mm}{
\begin{fmfgraph*}(12,0)
\fmfpen{thick}
\fmfleft{l}
\fmfright{r}
\fmfforce{(0.0w,0.5h)}{l}
\fmfforce{(1.0w,0.5h)}{r}
\fmf{plain,left=.9,tension=.5,label=$\pi$}{l,r}
\fmf{gluon,left=.1,label=$\rho$,l.d=7.5,l.side=right}{r,l}
\fmf{photon,left=.9,tension=.3,l.d=3.,label=$a_1$,l.side=right}{r,l}
\fmfsdot{l,r}
\end{fmfgraph*}
}}
\def\GPhipi{
\parbox{12mm}{
\begin{fmfgraph*}(12,0)
\fmfpen{thick}
\fmfleft{l}
\fmfright{r}
\fmfforce{(0.0w,0.5h)}{l}
\fmfforce{(1.0w,0.5h)}{r}
\fmf{plain,left=.9,tension=.5,l.d=1.5,label=$\pi$,l.side=right}{r,l}
\fmf{plain,left=.3,tension=.5,l.d=1.8,label=$\pi$,l.side=right}{r,l}
\fmf{plain,left=.3,tension=.5,l.d=1.7,label=$\pi$}{l,r}
\fmf{plain,left=.9,tension=.5,l.d=1.3,label=$\pi$}{l,r}
\fmfsdot{l,r}
\end{fmfgraph*}}
}
\def\GPipipi{
\parbox{20mm}{
\begin{fmfgraph*}(20,10)
\fmfpen{thick}
\fmfleft{l}
\fmfright{r}
\fmfforce{(0.0w,0.5h)}{l}
\fmfforce{(1.0w,0.5h)}{r}
\fmfforce{(0.2w,0.5h)}{ol}
\fmfforce{(0.8w,0.5h)}{or}
\fmf{plain,left=.7,label=$\pi$,l.s=right}{or,ol}
\fmf{plain,left=.7,label=$\pi$,l.s=right}{ol,or}
\fmf{gluon}{l,ol}
\fmf{gluon}{or,r}
\fmfsdot{ol,or}
\end{fmfgraph*}
}}
\def\GPipia{
\parbox{20mm}{
\begin{fmfgraph*}(20,10)
\fmfpen{thick}
\fmfleft{l}
\fmfright{r}
\fmfforce{(0.0w,0.5h)}{l}
\fmfforce{(1.0w,0.5h)}{r}
\fmfforce{(0.2w,0.5h)}{ol}
\fmfforce{(0.8w,0.5h)}{or}
\fmf{photon,left=.9,tension=.5,l.d=3.5,label=$a_1$,l.s=right}{or,ol}
\fmf{plain,left=.7,tension=.5,label=$\pi$,l.s=right}{ol,or}
\fmf{gluon}{l,ol}
\fmf{gluon}{or,r}
\fmfsdot{ol,or}
\end{fmfgraph*}
}}
\def\GPipirho{
\parbox{20mm}{
\begin{fmfgraph*}(20,10)
\fmfpen{thick}
\fmfleft{l}
\fmfright{r}
\fmfforce{(0.0w,0.5h)}{l}
\fmfforce{(1.0w,0.5h)}{r}
\fmfforce{(0.2w,0.5h)}{ol}
\fmfforce{(0.8w,0.5h)}{or}
\fmf{gluon,right=.5,label=$\rho$,l.s=left,l.dist=3}{ol,or}
\fmf{plain,left=.8,label=$\pi$,l.s=right,l.dist=3}{ol,or}
\fmf{photon}{l,ol}
\fmf{photon}{or,r}
\fmfsdot{ol,or}
\end{fmfgraph*}
}}
\def\GSigpirho{
\parbox{20mm}{
\begin{fmfgraph*}(20,10)
\fmfpen{thick}
\fmfleft{l}
\fmfright{r}
\fmfforce{(0.0w,0.5h)}{l}
\fmfforce{(1.0w,0.5h)}{r}
\fmfforce{(0.2w,0.5h)}{ol}
\fmfforce{(0.8w,0.5h)}{or}
\fmf{gluon,right=.5,label=$\rho$,l.side=left,l.dist=3}{ol,or}
\fmf{plain,right=.8,label=$\pi$,l.side=left,l.dist=3}{or,ol}
\fmf{plain}{l,ol}
\fmf{plain}{or,r}
\fmfsdot{ol,or}
\end{fmfgraph*}
}}
\def\GSigarho{
\parbox{20mm}{
\begin{fmfgraph*}(20,10)
\fmfpen{thick}
\fmfleft{l}
\fmfright{r}
\fmfforce{(0.0w,0.5h)}{l}
\fmfforce{(1.0w,0.5h)}{r}
\fmfforce{(0.2w,0.5h)}{ol}
\fmfforce{(0.8w,0.5h)}{or}
\fmf{photon,left=0.9,label=$a_1$,l.side=right,l.dist=3}{ol,or}
\fmf{gluon,right=.5,label=$\rho$,l.side=left}{ol,or}
\fmf{plain}{l,ol}
\fmf{plain}{or,r}
\fmfsdot{ol,or}
\end{fmfgraph*}
}}
\def\GSigpi{
\parbox{20mm}{
\begin{fmfgraph*}(20,10)
\fmfpen{thick}
\fmfleft{l}
\fmfright{r}
\fmfforce{(0.0w,0.5h)}{l}
\fmfforce{(1.0w,0.5h)}{r}
\fmfforce{(0.2w,0.5h)}{ol}
\fmfforce{(0.8w,0.5h)}{or}
\fmf{plain,left=.7,label=$\pi$,l.s=right}{or,ol}
\fmf{plain,left=.7,label=$\pi$,l.s=left}{ol,or}
\fmf{plain,label=$\pi$}{or,ol}
\fmf{plain}{l,ol}
\fmf{plain}{or,r}
\fmfsdot{ol,or}
\end{fmfgraph*}
}}
\def\Gcross#1#2{\fmfiv{d.sh=cross,d.ang=#1,d.size=10thick}{#2}}
\def\Gjmu{\parbox{15mm}
{\begin{fmfgraph*}(14,15)
\fmfpen{thick}
\fmfright{r}
\fmfleft{i1}
\fmf{gluon,left=.1,label=$\rho$}{i1,v1}
\fmf{phantom,tension=0.4}{v1,r}
\fmffreeze
\fmf{plain,left,label=$\pi$}{v1,r}
\fmf{plain,left}{r,v1}
\fmfsdot{v1}
\end{fmfgraph*}
}}
\def\GPhijmu{\parbox{20mm}
{\begin{fmfgraph*}(20,15)
\fmfpen{thick}
\fmfright{r}
\fmfleft{i1}
\fmf{gluon,left=.1,label=$\rho$}{i1,v1}
\fmf{phantom}{v1,r}
\fmffreeze
\fmf{plain,left,label=$\pi$}{v1,r}
\fmf{plain,left}{r,v1}
\fmfsdot{v1}
\fmffreeze
\Gcross{0}{vloc(__i1)}
\end{fmfgraph*}
}}
\def\GSigjmu{\parbox{10mm}
{\begin{fmfgraph*}(10,15)
\fmfpen{thick}
\fmfright{r}
\fmfleft{l}
\fmftop{t}
\fmfbottom{b}
\fmf{plain}{l,v,r}
\fmf{gluon,left=.1,label=$\rho$}{v,t}
\fmf{phantom}{v,b}
\fmfsdot{v}
\Gcross{0}{vloc(__t)}
\end{fmfgraph*}
}}
\def\PiVert{\parbox{20mm}
{\begin{fmfgraph*}(20,7)
\fmfpen{thick}
\fmfstraight
\fmfright{r}
\fmfleft{l}
\fmftop{t1,t2,t3}
\fmfbottom{b1,b2,b3}
\fmf{gluon,left=.1,tension=1,label=$\rho$}{v,r}
\fmfpoly{shaded,tension=0.7}{v,t2,b2}
\fmf{plain}{t2,vl,b2}
\fmf{gluon,left=.1,tension=3,label=$\rho$}{l,vl}
\fmfsdot{v,vl,t2,b2}
\end{fmfgraph*}
}}
\def\GVert{\parbox{17mm}
{\begin{fmfgraph*}(17,10)
\fmfpen{thick}
\fmfstraight
\fmfright{r1,r2,r3,r,r4,r5,r6}
\fmfleft{l}
\fmftop{t1,t2,t3,t4}
\fmfbottom{b1,b2,b3,b4}
\fmf{phantom}{t1,vt}
\fmf{phantom,tension=0.7}{vt,r4}
\fmf{phantom}{b1,vb}
\fmf{phantom,tension=0.7}{vb,r3}
\fmffreeze
\fmf{gluon,left=.1,tension=0.2,label=$\rho$}{v,r}
\fmfpoly{shaded,tension=0.7}{v,vt,vb}
\fmf{plain}{t1,vt}
\fmf{plain}{b1,vb}
\fmfsdot{v,t1,vt,b1,vb}
\end{fmfgraph*}
}}
\def\GKVert{\parbox{8mm}
{\begin{fmfgraph*}(8,8)
\fmfpen{thick}
\fmftop{t0,t1,t2}
\fmfbottom{b0,b1,b2}
\fmf{plain}{t0,t1,t2}
\fmf{plain}{b0,b1,b2}
\fmf{plain,left=.4}{t1,b1,t1}
\fmfsdot{t1,b1}
\end{fmfgraph*}
}}
\def\GVertLp{\parbox{24mm}
{\begin{fmfgraph*}(24,10)
\fmfpen{thick}
\fmfstraight
\fmfright{r1,r2,r3,r,r4,r5,r6}
\fmfleft{l}
\fmftop{t0,t1,t2,t3,t4}
\fmfbottom{b0,b1,b2,b3,b4}
\fmf{phantom}{t0,v1,vt}
\fmf{phantom}{t1,v1}
\fmf{phantom}{b1,v2}
\fmf{phantom,tension=0.7}{vt,r4}
\fmf{phantom}{b0,v2,vb}
\fmf{phantom,tension=0.7}{vb,r3}
\fmffreeze
\fmf{gluon,left=.1,tension=0.2,label=$\rho$}{v,r}
\fmfpoly{shaded,tension=0.7}{v,vt,vb}
\fmf{plain}{t0,v1,vt}
\fmf{plain}{b0,v2,vb}
\fmf{plain,left=.4}{v1,v2,v1}
\fmfsdot{t0,b0,v,v1,vt,v2,vb}
\end{fmfgraph*}
}}
\def\GVertV{\parbox{10mm}
{\begin{fmfgraph*}(10,10)
\fmfpen{thick}
\fmfstraight
\fmfleft{l1,l2}
\fmfright{r}
\fmf{plain,tension=0.35}{l1,v,l2}
\fmf{gluon,left=.1,label=$\rho$}{v,r}
\fmfsdot{l1,v,l2}
\end{fmfgraph*}
}}

\unitlength=1mm
\begin{fmffile}{kbd}
\fmfset{thick}{1.5pt}
\begin{equation}\label{Phi_Dyson}
\begin{array}{rccccc}
\Phi=&\displaystyle\frac{1}{2}\;\; \GPhipirhopi\;\;\;\;
&+&\displaystyle\frac{1}{2}\;\; \GPhipirhoa\;\;\;\;
&+&\displaystyle\frac{1}{2}\;\; \GPhipi\;\;\;\; \\[0.8cm] 
\Pi_{\rho}=&\GPipipi &+&\GPipia \\[0.8cm]
\Pi_{a_1}=&&&\GPipirho \\[0.8cm]
\Sigma_{\pi}=&\GSigpirho &+&\GSigarho &+&\GSigpi 
\end{array}
\end{equation}

\begin{figure}
\begin{picture}(140,60)
\put(-6,60){
\includegraphics[width=6cm,height=7.4cm,angle=-90]{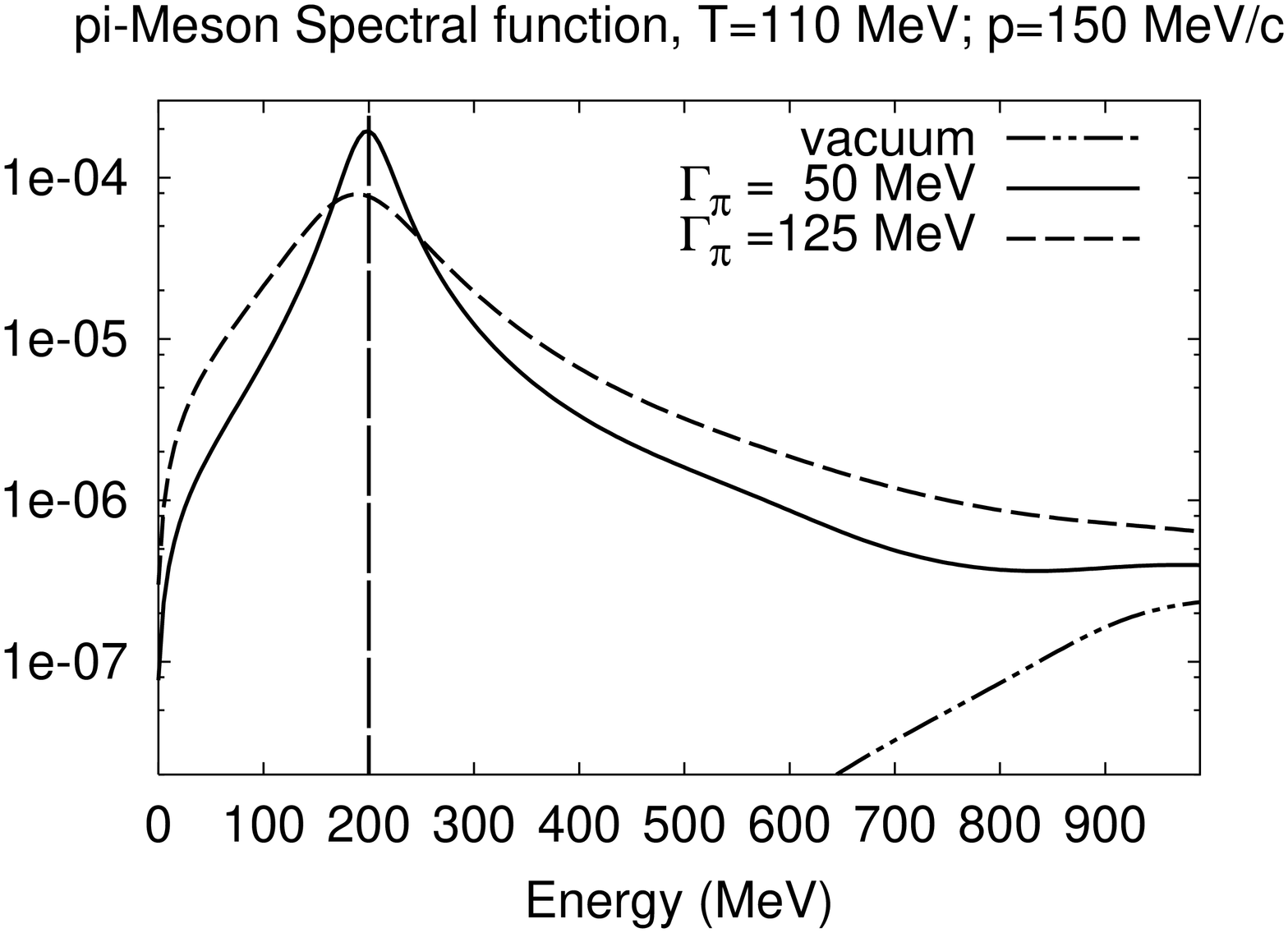}}
\put(68,60){
\includegraphics[width=6cm,height=7.4cm,angle=-90]{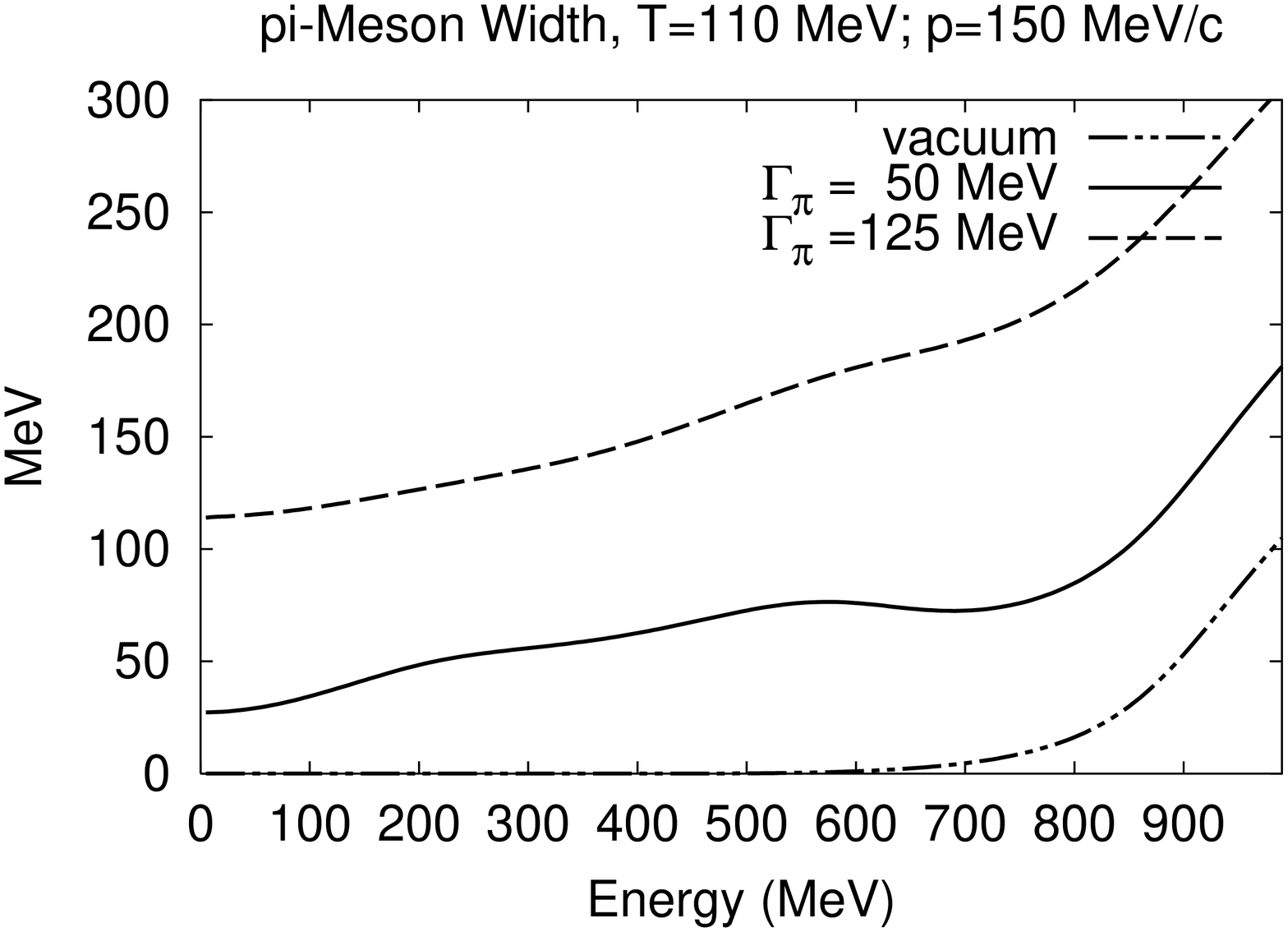}}
\end{picture}
\caption{\textit{Spectral function (left) and decay width (right) of the pion
    as a function of the pion energy at a pion momentum of $150\;
    \text{MeV}/c$ in the vacuum and for two self--consistent cases
    discussed in the text.}}
\label{fig1}
\end{figure}

They are the driving terms for the corresponding three Dyson
equations, which form a coupled scheme which has to be solved
self--consistently. The $\Phi$--derivable scheme pictorially
illustrates the concept of Newton's principle of \emph{actio =
  reactio} and detailed balance. If the self--energy of one particle is
modified due to the coupling to other species, these other species
also obtain a corresponding term in their self--energy.  In the vacuum
the $\rho$-- and $a_1$--meson self--energies have the standard
thresholds at $\sqrt{s}=2m_{\pi}$ and at $3 m_{\pi}$ respectively. For
the pion as the only stable particle in the vacuum with a pole at
$m_{\pi}$ the threshold opens at $\sqrt{s}=3 m_{\pi}$ due to the first
and last diagram of $\Sigma_{\pi}$.  Correspondingly the vacuum
spectral function of the pion shows already spectral strength for
$\sqrt{s}>3 m_{\pi}$, c.f. fig.~\ref{fig1} (left).

Self--consistent equilibrium calculations are performed keeping the
full dependence of all two--point functions on three momentum ${\vec
  p}$ and energy $p_0$, and treating all propagators with their
dynamically determined self--energies.

The examples shown refer to a temperature of $T=110 \,\text{MeV}$ appropriate
for the CERES data. We discuss three different settings.  First the
$\rho$--meson polarisation tensor is calculated simply by the perturbative
pion loop, i.e. with vacuum pion propagators and thermal Bose--Einstein
weights (no self--consistent treatment). The two other cases refer to
self--consistent solutions of the coupled Dyson scheme, where the four--$\pi$
interaction is tuned such that the sun--set diagram provides a moderate pion
damping width of about $50\,\text{MeV}$ and a strong one of $125\,\text{MeV}$
around the peak of the pion spectral function, c.f.  fig.~\ref{fig1}. Since in
the thermal case any excitation energy is available, though with corresponding
thermal weights, all thresholds disappear and the spectral functions show
strength at \emph{all} energies\footnote{In mathematical terms: all
branch--cuts in the complex energy plane reach from $-\infty$ to $+\infty$,
and the physical sheets of the retarded functions are completely separated
from the physical sheets of the corresponding advanced functions by these
cuts.}!  The pion functions shown in Fig.~\ref{fig1} at a fixed momentum of
$150 \text{MeV}$ are plotted against energy in order to illustrate that there
is significant strength also in the space--like region (below the light cone
at $150\,\text{MeV}$) resulting from $\pi$--$\pi$ scattering processes.

As an illustration we display a 3--d plot of the $\rho$--meson spectral
function as a function of $p_0$ and ${|\vec p\,|}$ in Fig.~\ref{fig2}, top
left.  The right part shows the transverse spectral function as a function of
invariant mass at fixed three--momentum of $150\,\text{MeV}/c$ in vacuum and
for the two self--consistent cases. The minor changes at the low mass
side of the $\rho$-meson spectral function become significant in the dilepton
yields given in the left bottom panel. The reason lies in the statistical
weights together with additional kinematical factors $\propto m^{-3}$ from the
dilepton--decay mechanism described by the vector meson dominance
principle\cite{klz67}. For the moderate damping case
($\Gamma_{\pi}=50\,\text{MeV}$) we have decomposed the dilepton rate into
partial contributions associated with $\pi$--$\pi$ bremsstrahlung,
$\pi$--$\pi$ annihilation and the contribution from the $a_1$--meson, which
can be interpreted as the $a_1$ Dalitz decay.

\unitlength=1mm
\begin{figure}
\begin{picture}(118,120)
\put(0,65){
\put(-8,74){{
\includegraphics[width=8cm,height=8cm,angle=-90]{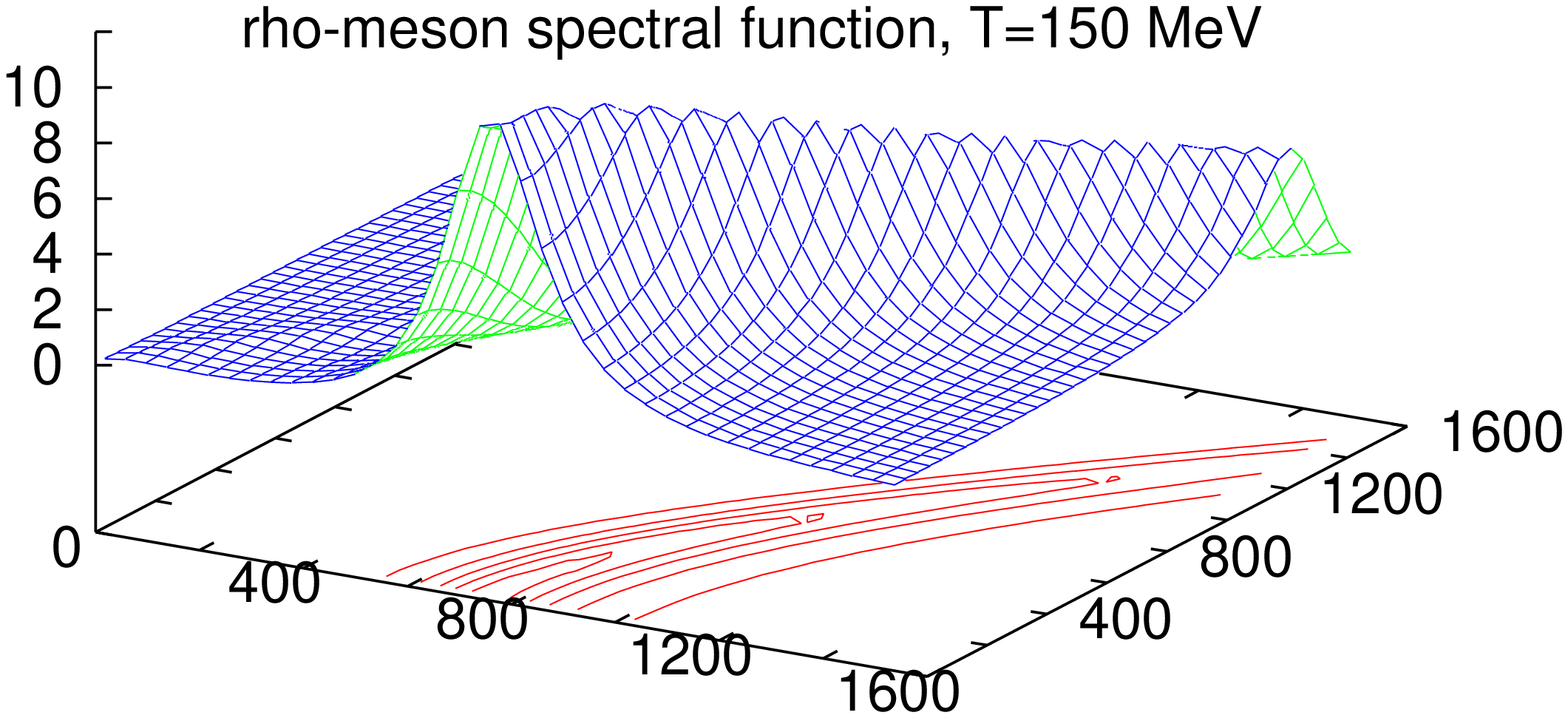}}}
\put(18,5){$p_0$}
\put(60,12){$|\vec p\,|$}
\put(63,55){{
\includegraphics[width=6cm,height=8cm,angle=-90]{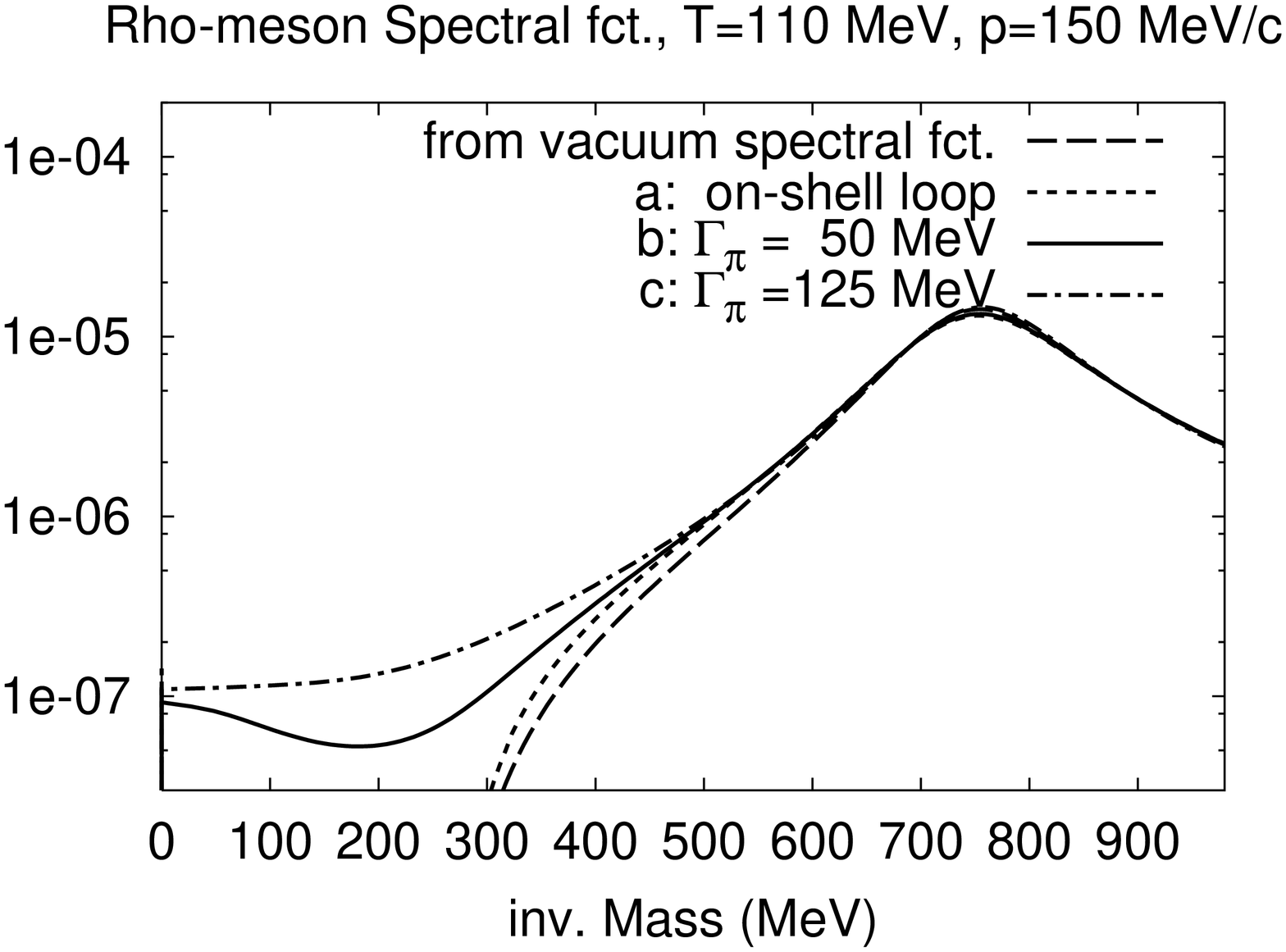}}}
}
\put(-6,60){{
\includegraphics[width=6cm,height=7.8cm,angle=-90]{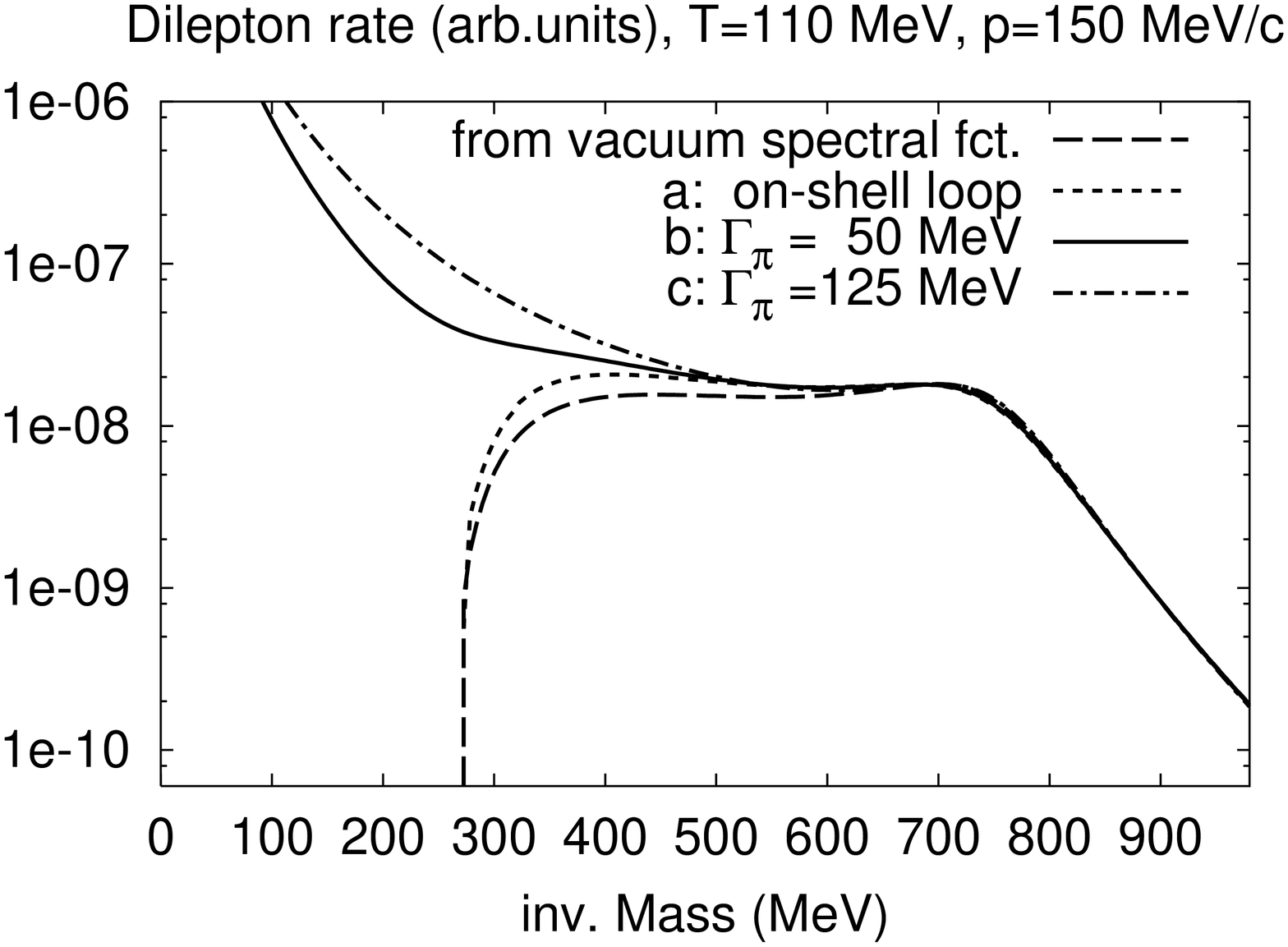}}}
\put(65,60){{
\includegraphics[width=6cm,height=7.8cm,angle=-90]{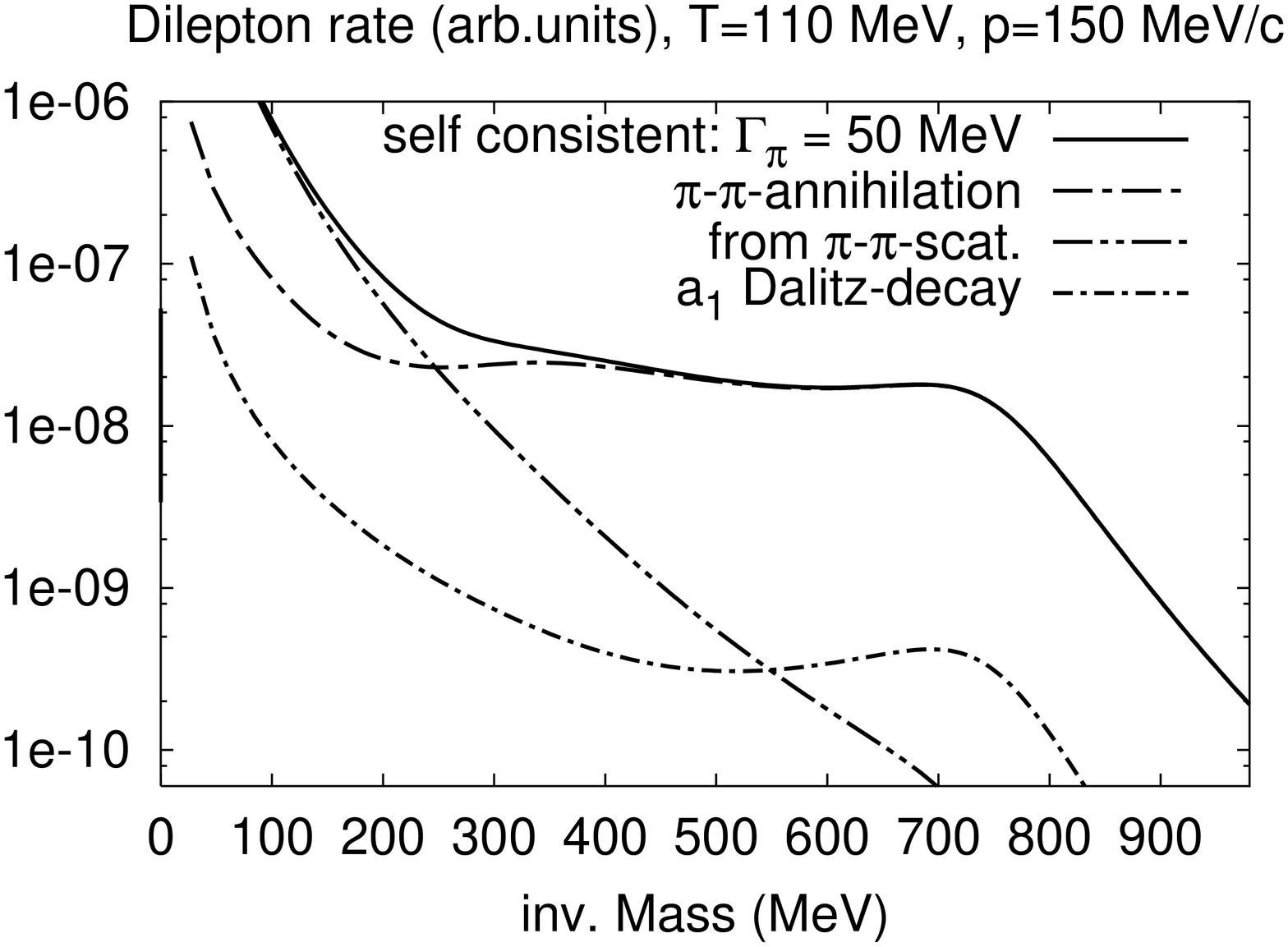}}}
\end{picture}
\caption{\textit{top: $\rho$--meson spectral function, bottom: thermal
    dilepton rate.}}
\label{fig2}
\end{figure}

The low mass part is completely dominated by pion bremsstrahlung contributions
(like--charge states in the pion loop). This contribution, which vanishes in
lowest order perturbation theory, is \emph{finite} for pions with finite width.
It has to be interpreted as bremsstrahlung, since the finite width results
from collisions with other particles present in the heat bath. Compared to the
standard treatment, where the bremsstrahlung is calculated independently of
the $\pi$--$\pi$ annihilation process, this self--consistent treatment has a
few advantages. The bremsstrahlung contribution is calculated consistently
with the annihilation process, it appropriately accounts for the
Landau--Pomeranchuk suppression at low invariant masses \cite{knoll96} and at
the same time includes the in--medium pion electromagnetic form--factor for
the bremsstrahlung part. As a result the finite pion width adds significant
strength to the mass region below $500\,\text{MeV}$ compared to the trivial
treatment with the vacuum spectral function. Therefore the resulting dilepton
spectrum essentially shows no drop any more in this low mass region already
for a moderate pion width of $50 \, \text{MeV}$. The $a_1$ Dalitz decay
contribution can be read off from the partial $\rho$--meson width due to the
$\pi$--$a_1$ loop in $\Pi_{\rho}$. This component is seen to be unimportant at
all energies in the present calculations where medium modifications of the
masses of the mesons are discarded. The latter can be included through
renormalised dispersion relations within such a consistent scheme.

\section{Longitudinal and transverse components}

While scalar particles and couplings can be treated self--consistently with no
principle problems at any truncation level, considerable difficulties and
undesired features arise in the case of vector particles. The origin lies in
the fact that, though in $\Phi$--derivable Dyson resummations symmetries and
conservation laws are fulfilled at the {\emph expectation} value level, they
are generally no longer guaranteed at the correlator level. Considering the
$\rho$-meson as a gauge particle one has to care about local gauge symmetries,
where the situation is even worse, because the symmetry of the quantised
theory is not the original one but the non--linear BRST symmetry
\cite{brs76,tyfr72}. Contrary to perturbation theory, where the loop expansion
corresponds to a strict power expansion in $\hbar$ and symmetries are
maintained order by order, partial resummations mix different orders and thus
are violating the corresponding symmetries.  It is obvious that the scheme
discussed above indeed violates the Ward identities on the correlator level
and thus the vector--meson polarisation tensor is no longer 4--dimensionally
transverse.  This means that unphysical states are propagated within the
internal lines of the $\Phi$--derivable approximation scheme which leads to a
number of conceptual difficulties and to explicit difficulties in the
numerical treatment of the problem. In the above calculations we have worked
around this problem in the following way.

For the exact polarisation tensor we know that for $\vec{p}=0$ the temporal
components exactly vanish $\Pi^{00}(q)=\Pi^{0i}(q)=\Pi^{i0}(q)=0$ for $q_0\ne
0$, while this is not the case for the self--consistently constructed tensor
\cite{knoll96}. Indeed these components are tied to the conservation of charge
and therefore involve a relaxation time for a conserved quantity which is of
course infinite while the self--consistent result always reflects the damping
time of the propagators in the loop. This behaviour is studied in detail in
ref. \cite{knoll96}, both on the classical and quantum many body Green's
function level within the real time formalism. There it has been shown that
current conservation can only be restored through a resummation of all the
scattering processes in a transport picture which amounts to a Bethe--Salpeter
ladder resummation in the corresponding quantum field theory description. This
will be discussed to some extent in sect. 4. At this level it is important to
realise that the spatial components generally suffer less corrections from
this resummation in case that the relaxation time for the transverse
current--current correlator is comparable to the damping time of the
propagators in the loop. The time components, however, suffer significant
corrections. Thus our strategy for the self--consistent loop calculation is
the following: from the loop calculation of the polarisation tensors $\Pi$ (of
the $\rho$--meson) we evaluate only the information obtained for the spatial
components $\Pi^{ik}$.  Taking the following two spatial traces (details are
given in the appendix A)
\begin{eqnarray}
\left.
\begin{array}{rcccl}
3 \Pi_3&=&
\displaystyle
-g_{ik}\Pi^{ik}&=&2\Pi_T
+\frac{(q^0)^2}{q^2}\Pi_L
\\
\displaystyle
\Pi_1&=&\displaystyle
\frac{p_i p_k}{{\vec q}^2}\Pi^{ik}&=&
\displaystyle
\frac{(q^0)^2}{q^2}\Pi_L
\end{array}
\right\}
\quad\Leftarrow\quad \partial_{\mu}\Pi^{\mu\nu}=0
\end{eqnarray}
permits to deduce the 3--dim. longitudinal and transverse tensor
components $\Pi_L$ and $\Pi_T$ under the condition that the polarisation
tensor is exactly 4--dim. transversal. This construction thus fulfils current
conservation on the correlator level.

\begin{figure}
\begin{picture}(150,75)
\put(-5,0){
\put(0,75){
\includegraphics[width=7.5cm,height=6.8cm,angle=-90]{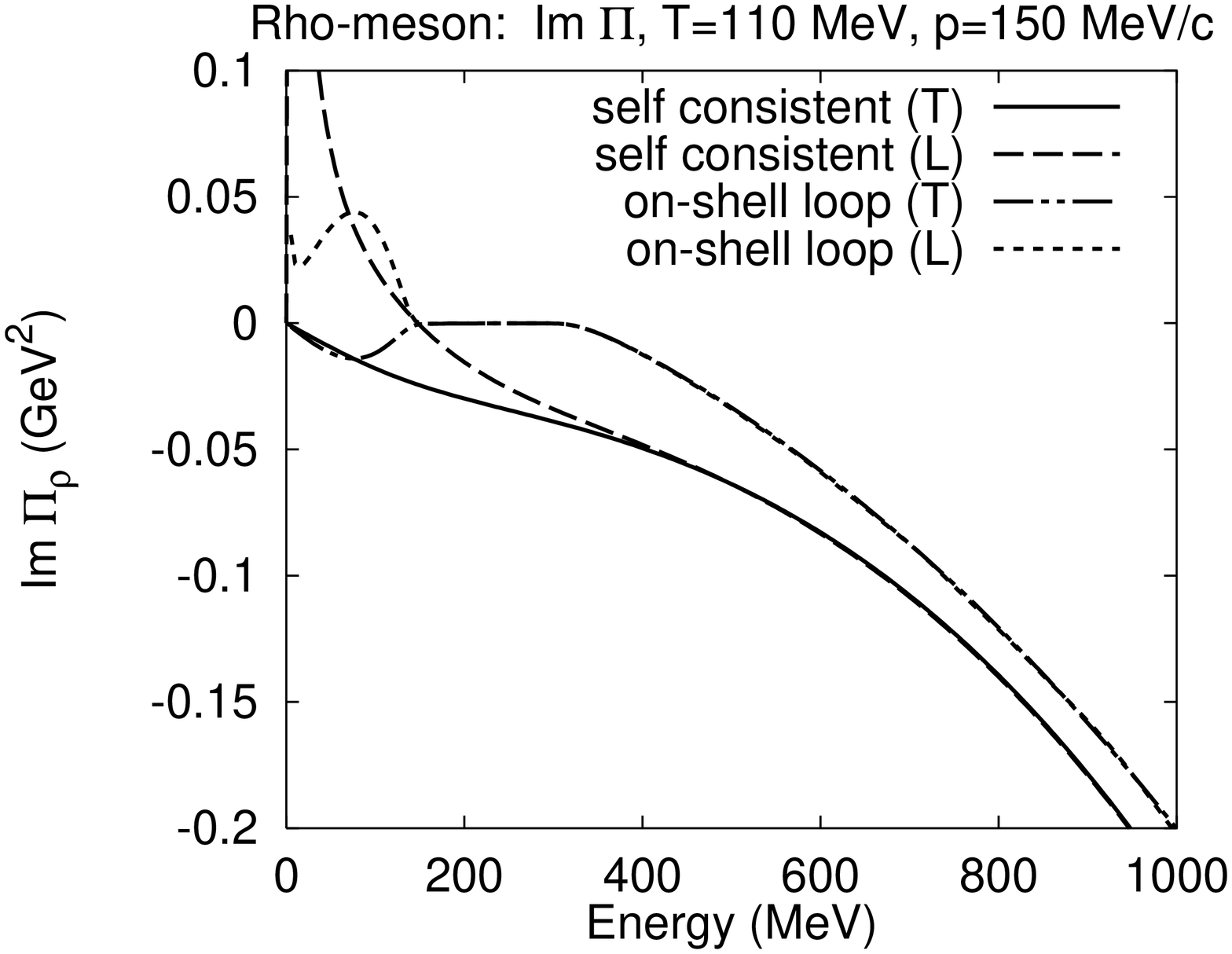}}
\put(48,40){a}\put(48,27){b}
\put(75,75){
\includegraphics[width=7.5cm,height=6.8cm,angle=-90]{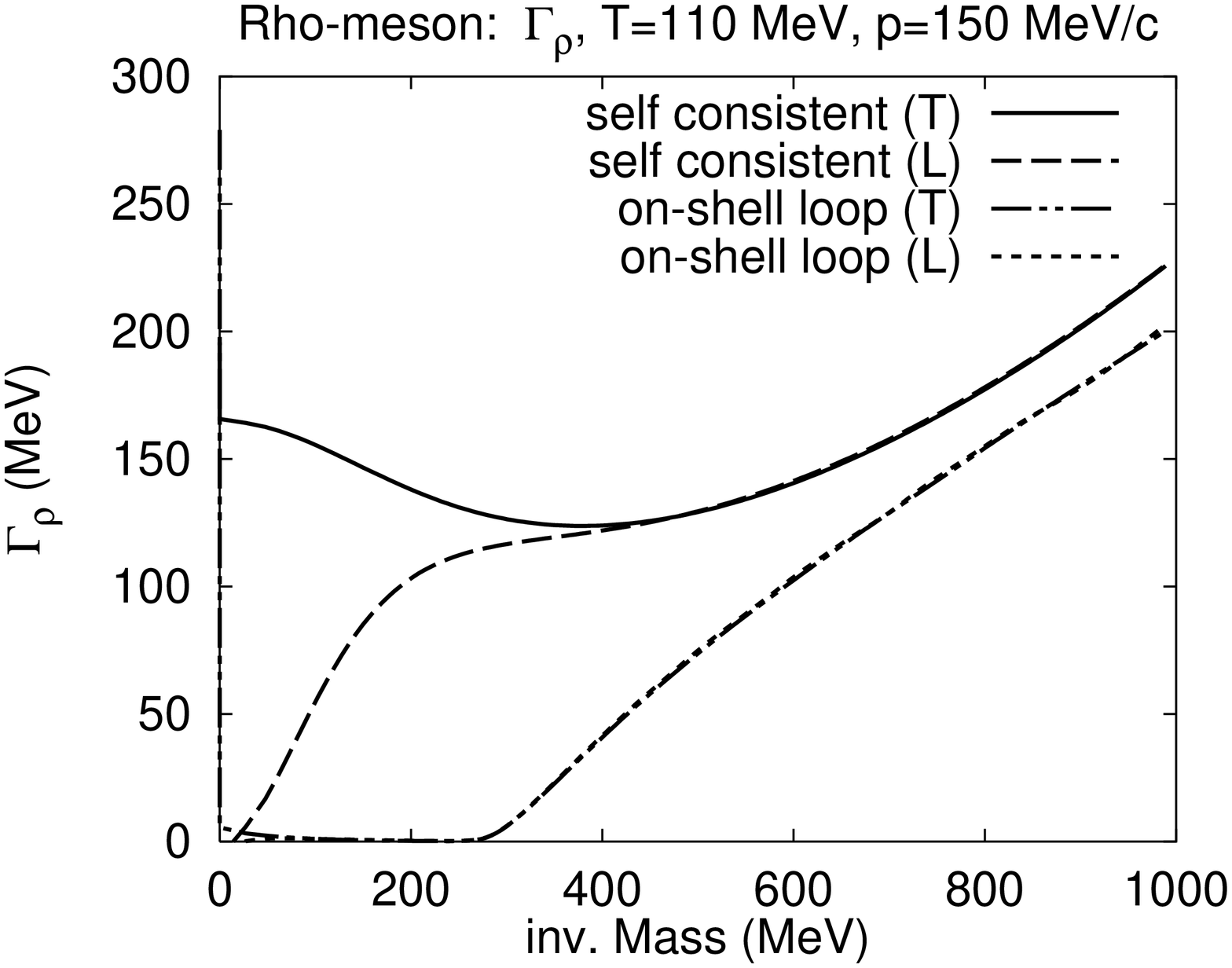}}
\put(116,21){a}\put(116,37){b}
}
\end{picture}
\caption{\textit{Left: Imaginary part of the Longitudinal (L) and transverse
    (T) $\rho$--meson polarisation tensor at $T=110 \,\text{MeV}$.
    case (a): finite temperature on-shell--loop calculation; case (b):
    self--consistent calculation for the case where $\Gamma_{\pi}=125
    \, \text{MeV}$. Right: the same for the corresponding widths
    $\Gamma_{\rho}=-\Im \Pi_{\rho}/p^0$.}}
\label{Pi_L_T}
\end{figure}

The result of this procedure is shown in fig.~\ref{Pi_L_T} for the
components of the $\rho$--meson polarisation tensor for a finite
spatial momentum of $\vec{p}=150\,\text{MeV}$. The plots show $\Im
\Pi_L$ and $\Im \Pi_T$ first for the on--shell loop result, i.e. with
vacuum pion spectral functions and thermal occupations. For this
on--shell loop case to very good approximation one finds $\Pi_L=\Pi_T$
for time like momenta, while as expected they deviate in sign for
space--like momenta. The longitudinal component exactly vanishes on
the light cone and changes sign there. Thus the tensor is entirely
transverse on the light cone as it should! Switching to the self--consistent
results the threshold gap between $\sqrt{s}\in [0,2m_{\pi}]$ is
completely filled. At non--zero momenta $\vec{p}$ the longitudinal and
tranverse component deviate from one another towards low invariant
masses, i.e. $\sqrt{s}< 400 \, \text{MeV}$ in this case, while they
are identical for large $\sqrt{s}$ as they should. As both components
$\Pi_L$ and $\Pi_T$ are constructed from different moments of the
numerically given $\Pi^{ik}$, the agreement of the two components at
large $\sqrt{s}$ shows the numerical precision of the employed loop
integration method.  The resulting behaviour is further clarified in
the right part of fig.~\ref{Pi_L_T}, which shows the resulting damping
width of the $\rho$--meson $\Gamma_{\rho}(p)=-\Im \Pi_{\rho}(p)/p^0$. One
sees that the typical threshold behaviour of the on--shell loop is
completely changed in the self--consistent result. The transverse
width is with $\Gamma_T\approx 150 \text{MeV}$ almost constant over
the displayed invariant mass range! For the longitudinal component one
has to consider the kinematical factor entering in specific tensor
components, e.g. $\Pi^{00}=\Pi_L\;{\vec q}^2/q^2$, such that also
$\Gamma^{00}$ is about constant.

While current conservation has been restored on the correlator level by the
procedure above, the Ward--Takahashi identities are certainly not fulfilled.
Thus the whole procedure will not be gauge covariant.  Still from the
experience discussed in ref.~\cite{knoll96} we expect that this method
provides a good approximation to the in--medium polarisation tensor at finite
temperature. In particular for the light pions which at $T=150\, \text{MeV}$
have already quite relativistic energies we expect that the mutual scattering
leads to fairly isotropic distributions after each scattering such that the
memory on some initial fluctuation of the pion current is already lost after
the first collision, c.f. the discussions in ref. \cite{knoll96} and in the
next section.

\section{Symmetries and gauge invariance}

In view of the difficulties to provide a gauge--invariant scheme one
may raise the question: \emph{is there a self--consistent truncation
  scheme beyond the mean field level for the gauge fields, which
  preserves gauge invariance?} In particular we are interested that
the internal dynamics, i.e. the dynamical quantities like classical
fields and propagators which enter the self--consistent set of
equations remain gauge covariant.

At the mean field level the gauge fields couple to the expectation values of
the vector currents and gauge covariance is fully maintained. This level is
explored in all hard thermal loop (HTL) approaches
\cite{Braaten91,Pisarski91,Blaizot93,Jackiw93}. For the 
$\pi$--$\rho$--meson system the mean field
approximation is given by the following $\Phi$--derivable scheme (again
omitting the tadpole term for the pion self--energy)
\begin{eqnarray}
\label{piextrho}
\Phi\{G_{\pi},\rho\}&=&\hspace*{2mm}\GPhijmu\;\,+\hspace*{5mm}\GPhipi \\
\label{Sig-pi-cl}
\Sigma_{\pi}&=&\hspace*{7mm}\GSigjmu\hspace*{7mm}+\;\GSigpi\\
\label{cl-field-eq}
\left(\partial^{\nu}\partial_{\nu}-m^2\right) \rho^{\mu}&=&
j^{\mu}=\;\Gjmu
\end{eqnarray}
Here full lines represent the self--consistent pion propagators and
curly lines with a cross represent the classical $\rho$--meson field,
governed by the classical field equations of motion
(\ref{cl-field-eq}). Since $\Phi$ is invariant with respect to gauge
transformations of the classical vector field $\rho^{\mu}$, the
resulting equations of motion are gauge covariant.

The step to construct symmetry preserving correlation functions is provided by
considering the linear response of the system on fluctuations in the
background field \cite{bk61,baym62}, see also \cite{den96} in the context of
gauge and Goldstone bosons. Thereby gauge covariance also holds for
fluctuations $\rho^{\mu}+\delta \rho^{\mu}$ around mean field solution of
(\ref{Sig-pi-cl} - \ref{cl-field-eq}). Thus, one can then define a gauge
covariant \emph{external} polarisation tensor via variations with respect to
the background field $\delta \rho^{\mu}$
\begin{equation}
\label{Pi-ladder}
\Pi^{\rm ext}_{\mu \nu}(x_1,x_2)=
\frac{\delta}{\delta \rho^{\mu}(x_2)}
\left.\frac{\delta \Phi[G_{\pi},\rho]}{\delta  \rho^{\nu}(x_1)}
\right|_{G_{\pi}[\rho]} =\; \PiVert
\end{equation}
as a linear response to fluctuations around the mean field. This
tensor can be accessed through a corresponding three--point--vertex
equation
\begin{equation}
\label{vert-ladder}
\funcd{G_{\pi}}{\rho^{\mu}}=\;\GVert\;=\;\GVertV\;+\;\;\GVertLp
\end{equation}
In order to maintain all symmetries and invariances, the four-point
Bethe--Salpeter Kernel in this equation has to be chosen consistently
with the $\Phi$--functional (\ref{piextrho}) \cite{baym62,bk61}, i.e.
as a second functional variation of $\Phi$ with respect to the
propagators
\begin{equation}
\label{ladder-kernel}
K_{1234}=\frac{\delta^2\Phi}{\delta G_{12}\delta G_{34}}.
=\;\GKVert
\end{equation}
Thereby the pion propagator entering the ladder resummation
(\ref{vert-ladder}) is determined by the self--consistent solution of
Eq. (\ref{Sig-pi-cl}) at vanishing classical $\rho$-field. In
particular this ladder resummation accounts for real physical
scattering processes, a phenomenon already discussed in \cite{knoll96}
for the description of Brems\-strahlung within a classical transport
scheme (Landau--Pomeranchuk--Migdal effect). From this point of view
one clearly sees that the pure $\Phi$--functional formalism without
the vertex corrections (\ref{vert-ladder}) just describes the ``decay
of states'' due to collision broadening. Thus in the $\Phi$--Dyson
scheme (\ref{Phi_Dyson}) all components of the \emph{internal}
$\rho$--meson polarisation tensor have a \emph{time--decaying}
behaviour with a decay constant given by the pion damping rate
$\Gamma_{\pi}$. However, the exact tensor has at least two decay
times, one for the transverse components and a second one which
involves the conserved charge and which naturally is infinite. The
Dyson resummation fails to cope with this, since there also the
$00$--component approximately behaves like
\begin{equation}
\label{Pi00-time1}
\Pi^{00}_{\rho}(\tau,{\vec p}=0)\propto e^{-\Gamma_{\pi} \tau}
\end{equation}
in a mixed time--momentum representation. This clearly violates charge
conservation, since $\partial_0\Pi^{00}_{\rho}(\tau,{\vec p})|_{{\vec
    p}=0}$ does not vanish! Yet, accounting coherently for the
multiple scattering of the particles through the vertex resummation
(\ref{vert-ladder}) keeps track of the ``charge flow'' into other
states and thus restores charge conservation. Within classical
considerations the ladder resummation
(\ref{vert-ladder}) indeed yields
\begin{equation}
\label{Pi00-time2}
\Pi^{00}_{\rho}(\tau,{\vec p}=0)\propto 
\sum_n \frac{(\Gamma\tau)^n}{n!}\; e^{-\Gamma \tau}=1
\end{equation}
confirming charge conservation. For further details c.f. ref.
\cite{knoll96}. From the physics discussions above it is clear that
these conclusions hold also for constant self--energies with a constant
imaginary part. This is opposed to the vacuum case where a constant
self--energy would not require any vertex correction! The formal origin
of this difference lies in the fact that in the real time formulation
of the field theory all relations become matrix relations from the
contour time ordering. In particular the three point functions then
have three independent retarded components\footnote{In the Matsubara
  formalism this corresponds to the fact that the different energy
  arguments of three point functions can be placed on different half
  planes.} (c.f. \cite{heinz98,wang99}) and the corresponding
Ward-Takahashi identities involve both retarded and advanced self
energy terms which differ in the sign of their imaginary parts. Thus
even for constant self--energies the terms related to the width do no
longer cancel out in these identities in the true real time case at
finite $T$.

\section{Comments and prospects}

We presented self--consistent calculations of the vector meson
production in a pion gas environment, where the width of the pion was
generated by a pion four-vertex interaction. The novel part is that
through the damping width standard thresholds known from vacuum
calculations disappear and that contributions arising from pion
bremsstrahlung and from $\pi$--$\pi$--annihilation are treated within
the same scheme, and are therefore consistent with one another. With
reasonable damping width for the pions the calculations show
significant contributions to the dilepton spectrum in the mass range
below $400\,\text{MeV}$. In the second part we discussed the
particular features related to the self--consistent treatment of
vector or gauge bosons.

Vertex corrections of various types were already considered in the
literature in the context of the $\rho$--meson. They dealt with the
case, where the pion couples to the nuclear sector, and vertex
corrections related to the $p$--wave coupling of the
$\pi$-$N$-$\Delta$--vertex where included on phenomenological grounds
using Landau--Migdal parameterisations for the $\Delta$-$N$
interaction \cite{korpa90,asakawa92}. There the employed
quasi-particle loops are straightforward and the ``bubble''
resummation (algebraic) also. More involved vertex corrections were
considered in refs. \cite{hfn93} and by many others later, e.g.
\cite{ubw99}. However in none of those papers the vertex corrections
were done self--consistently, nor were they proven to restore gauge
invariance, nor has there the question been addressed which vertex
corrections are required, once the propagators in the self--energy
loop of the gauge particles have a significant damping width. The
latter question was addressed in ref.  \cite{knoll96} in the context
of photon production and put on formal grounds here. Using the
background field scheme we explained above the steps to come to a
consistent vertex equation (\ref{vert-ladder}-\ref{ladder-kernel}). It
is neither some vertex equation nor the exact vertex equation: it is
precisely \emph{the} vertex equation which pertains to the
self--consistent pion self--energy given by (\ref{Sig-pi-cl}) at
vanishing background field.  The consistency comes about by the fact
that both, the pion self energy (\ref{Sig-pi-cl}) and the
Bethe--Salpeter kernel (\ref{ladder-kernel}), are generated from the
same $\Phi$-functional (\ref{piextrho}). The method is general and
applies to any kind of $\Phi$-functional supplemented by terms
coupling to background gauge fields.

Two problems are to be mentioned in this context, a practical and a
principle one.  The practical problem concerns the fact that there is
no feasible algorithm to calculate the external polarisation tensor
(\ref{Pi-ladder}). Already for our most simple example one has to
solve the ladder re--summation in the Bethe--Salpeter equation with
\emph{full off--shell momentum dependence} for the three--point vertex
given by (\ref{vert-ladder}). Even if one restricts oneself to the
simplifying case of vanishing $\rho$--meson three--momentum for this
vertex the numerical effort increases by about two orders of magnitude
(for the full momentum dependence a factor $10^5$) compared to the
presented numerical solution of self--consistent self--energies and is
thus out of reach in practice.

However there is yet a principle problem. What one constructs by the
vertex equation is the \emph{external} polarisation tensor of the
vector meson. It has all desired features. However, the corresponding
vector--meson propagator does not take part in the self--consistent
scheme. This so constructed external propagator is fine in all cases,
where the vector meson couples perturbatively to a source, e.g. like
the photon to the electromagnetic current of a source system. For the
case of the $\rho$--meson recoupling effects may already be of some
importance, and for sure for gluons in an interacting quark-gluon
plasma such recoupling effects are important, and the vector meson is
a sensible component in the self--consistent scheme. In such cases,
however, one sees that in self--consistent truncation schemes there is
generally a difference between the self--consistent internal
propagator and the external propagator constructed from the
Bethe--Salpeter ladder resummation. While the first violates
Ward--Takahashi identities, the latter fulfils them.

Therefore presently we see no obvious self--consistent scheme where
vector particles are treated dynamically beyond mean field, i.e. with
dynamical propagators, and which at the same time complies with gauge
invariance also for the \emph{internal} propagation, unless one solves
the exact theory. The work around presented in sect. 3 at least
guarantees that the polarisation tensor remains four--dimensional
transverse and thus no unphysical modes are appearing in the scheme.
The problem of renormalisation omitted here is investigated separately
using subtracted dispersion relations \cite{vHJK00}. Thus for vector
particles a fully self--consistent scheme with all the features of the
$\Phi$--functional, especially to ensure the consistency of dynamical
and thermodynamical properties of the calculated propagators together
with the conservation laws on both the expectation value and the
correlator level remains an open problem.

The self--consistent equilibrium calculations presented here also
serve the goal to gain experience about particles with broad damping
width aiming towards a transport scheme for particles beyond the
quasi--particle limit \cite{ikv99-2}, see also
\cite{eff99,Leupold99,Cassing99}.

\section*{Acknowledgements}

We like to thank G. Baym, B. Friman, Y. Ivanov, E. Kolomeitsev, and D.
Voskresensky for helpful discussions. Part of the formal
considerations concerning gauge invariance were developed during our
stay at the workshop ``Non--Equilibrium Non-Equilibrium Dynamics in
Quantum Field Theory'' at the National Institute of Nuclear Theory at
the University of Washington, especially in discussions with E.
Mottola.

\appendix

\section{Decomposition of the polarisation tensor}

In spherically symmetric systems the polarisation tensor $\Pi$ can be
decomposed into three components (4--longitudinal (l) and the two
4-transverse components, the longitudinal (L) and transverse (T) one)
\begin{eqnarray}
\Pi^{\mu\nu}&=&
\Pi^{\mu\nu}_l+\Pi^{\mu\nu}_L+\Pi^{\mu\nu}_T\\
\Pi^{\mu\nu}_l&=&-\frac{q^{\mu}q^{\nu}}{q^2}\Pi_l
\\\Pi^{\mu\nu}_L&=&
\left(-g^{\mu\nu}-\delta^{\mu\nu}
        +\frac{q^{\mu}q^{\nu}}{q^2}
        +\frac{{\vec q}^{\mu}{\vec q}^{\nu}}{{\vec q}^2}
\right)\Pi_L\\
\Pi^{\mu\nu}_T&=&
\left(\delta^{\mu\nu}-\frac{{\vec q}^{\mu}{\vec q}^{\nu}}{{\vec q}^2}
\right)\Pi_T.
\end{eqnarray}
Here $\delta^{\mu\nu}$ and ${\vec q}^{\mu}$ are defined such that the
time--components vanish. The spatial part and the $00$--component
become
\begin{eqnarray}
\Pi^{ik}&=&
\left(\delta^{ik}-\frac{{\vec q}^{i}{\vec q}^{k}}{{\vec q}^2}
\right)\Pi_T
+\frac{(q^0)^2}{q^2}\frac{q^iq^k}{{\vec q}^2}\Pi_L
-\frac{q^iq^k}{q^2}\Pi_l\\
\Pi^{00}&=&
\left(-g^{00}+\frac{(q^0)^2}{q^2}\right)\Pi_L-\frac{(q^0)^2}{q^2}\Pi_l.
\end{eqnarray}
In terms of the 4-- and 3--traces we define
\begin{eqnarray}
4 \Pi_4&=-\Tr _4\{\Pi^{\mu\nu}\}&=-g_{\mu\nu}\Pi^{\mu\nu}
=-\Pi^{00}+\Tr_3\{\Pi^{ik}\}\cr
&&=\Pi_l+\Pi_L+2\Pi_T\\
3 \Pi_3&=\Tr _3\{\Pi^{ik}\}&=-g_{ik}\Pi^{ik}=2\Pi_T
+\frac{(q^0)^2}{q^2}\Pi_L
-\frac{{\vec q}^2}{q^2}\Pi_l.
\end{eqnarray}
One further can use
\begin{eqnarray}
\Pi_1=\frac{p_i p_k}{{\vec q}^2}\Pi^{ik}&=&
\frac{(q^0)^2}{q^2}\Pi_L-\frac{{\vec q}^2}{q^2}\Pi_l.
\end{eqnarray}
Taking into account the information contained in the two traces
$\Pi_1$ and $\Pi_3$, and the condition for 4--dim. transversality,
$\Pi_l=0$, a conserved polarisation tensor can be constructed.

\end{fmffile}
\bibliographystyle{elsart-num}
\bibliography{refs}

\end{document}